\documentclass[epj]{svjour}
\usepackage{graphics,amsfonts,amsmath,amssymb,amstext,bm}
\usepackage{color}
\begin{document}
\title{Chiral anomalous processes in magnetospheres of pulsars and black holes}
\author{Eduard~V.~Gorbar\inst{1,2} \and Igor~A.~Shovkovy\inst{3,4}% etc
}                     % Do not remove
\institute{Department of Physics, Taras Shevchenko National Kyiv University, Kyiv, 01601, Ukraine \and Bogolyubov Institute for Theoretical Physics, Kyiv, 03143, Ukraine \and College of Integrative Sciences and Arts, Arizona State University, Mesa, Arizona 85212, USA \and Department of Physics, Arizona State University, Tempe, Arizona 85287, USA}
\date{Received: 10 May 2022 / Accepted: 10 July 2022}
\abstract{
We propose that chirally asymmetric plasma can be produced in the gap regions of the magnetospheres of pulsars and black holes. We show that, in the case of supermassive black holes situated in active galactic nuclei, the chiral charge density and the chiral chemical potential are very small and unlikely to have any observable effects. In contrast, the chiral asymmetry produced in the magnetospheres of magnetars can be substantial. It can trigger the chiral plasma instability that, in turn, can lead to observable phenomena in magnetars. In particular, the instability should trigger circularly polarized electromagnetic radiation in a wide window of frequencies, spanning from radio to near-infrared. As such, the produced chiral charge has the potential to affect some features of fast radio bursts.   
\PACS{~
     {~}{~}  %\and
     %{~}{black holes}
     } % end of PACS codes
} %end of abstract
\maketitle
\section{Introduction}

Over the last decade, investigations of the chiral anomalous phenomena in relativistic plasmas attracted much attention. The main applications were focused on the anomalous effects in the quark-gluon plasma produced in heavy-ion collisions \cite{Kharzeev:2007jp,Kharzeev:2010gr,Gorbar:2011ya,Burnier:2011bf,Bzdak:2012ia,Kharzeev:2013ffa,Tuchin:2014iua,Liao:2014ava,Miransky:2015ava,Huang:2015oca,Kharzeev:2015znc}, hot plasma in the early Universe \cite{Joyce:1997uy,Boyarsky:2011uy,Boyarsky:2012ex,Tashiro:2012mf,Tashiro:2013bxa,Tashiro:2013ita,Manuel:2015zpa,Hirono:2015rla}, and quasirelativistic electron plasma in Dirac/Weyl semimetals \cite{Armitage:2017cjs,Gorbar:2021ebc}. Since relativistic plasmas and strong electromagnetic fields are widespread in astrophysics, it is also sensible to explore the role of anomalous effects there. Guided by this general idea, several anomalous phenomena were proposed in the context of supernova physics \cite{Charbonneau:2009ax,Ohnishi:2014uea,Kaminski:2014jda,Dvornikov:2014uza,Sigl:2015xva,Dvornikov:2015iua,Yamamoto:2015gzz}. In this study, we extend the idea further by addressing possible implications of the chiral anomaly in magnetospheres of black holes and pulsars. 

Both compact stars and black holes are surrounded by magnetospheres made of relativistic plasma that host powerful relativistic jets composed of ionized matter accelerated to nearly the speed of light, e.g., the observed Lorentz factors in jets of active galactic nuclei can exceed $\gamma \sim 100$ \cite{Blandford:2018iot}. While operating on vastly different spatial and temporal scales, magnetospheres of both systems have fascinating and often surprisingly similar properties \cite{Beskin:2010iba}. Therefore, our considerations below may be applied to both types of systems. 

Of the two, however, only the magnetospheres of magnetars are more likely to host superstrong electromagnetic fields that can trigger chiral anomalous effects in earnest. Indeed, the magnetic fields near accreting black holes are estimated to be at most about $10^4~\mbox{G}$ \cite{Beskin:2010iba}, i.e., they are much weaker compared to the critical magnetic field $B_{c}\equiv m_e^2/e \approx  4.4 \times 10^{13}~\mbox{G}$. Accordingly, as we will show, the possibility of chiral charge production in the active galactic nuclei is completely negligible. On the other hand, the values of magnetic fields near magnetars can reach up to about $10^{15}~\mbox{G}$ or even slightly higher \cite{Turolla:2015mwa,Kaspi:2017fwg,Gourgouliatos:2018efn}. Therefore, the chiral anomalous processes in their magnetospheres can indeed play a profound role. Estimating the chiral charge production rate and exploring its implications for magnetar physics will be the main focus of the current study.

The properties of magnetospheres and the origin of jets produced by compact stars is an active research area \cite{vandenEijnden:2018moe,Nishikawa:2020rwe,Komissarov:2021}. However, here we do not address the details of the magnetosphere formation or the jet dynamics. Instead, we focus on the role of the chiral anomaly in the so-called gap regions of the magnetosphere. For our purposes, it is sufficient to know that the magnetosphere is strongly magnetized, whereas its exact particle composition is unimportant. It is also critical that the gap (vacuum) regions should form intermittently at various places in the magnetosphere \cite{Sturrock:1971zc,Ruderman:1975ju,Arons:1983aa,Cheng:1986qt}. After appearing, the gaps are refilled with a relativistic electron-positron plasma that is induced by a background flux of energetic photons traversing the magnetized vacuum. (Since the analogous production of proton-antiproton pairs is strongly suppressed, the proton fraction in the same regions should be negligible.) As we will argue, the quantum anomaly can produce a substantial chiral charge density in the corresponding relativistic electron-positron plasma. The chiral charge, in turn, can modify the properties of plasma and affect the observable signatures of magnetars. Note that a different possibility was proposed recently in Ref.~\cite{Prabhu:2021zve}, where the chiral anomaly was argued to lead to the production of QCD axions that may also produce observable effects.

As we will argue, the chiral charge production in the pulsar magnetospheres may explain some characteristics of the observed fast radio bursts. While numerous models were proposed in the literature, the underlying mechanism of the fast radio bursts remains puzzling. The corresponding transient radio pulses of millisecond duration release a huge amount of energy, often as much energy as the Sun over several days. Because of the large magnetic energy density in their magnetospheres, magnetars were identified as possible sources of the fast radio bursts \cite{Masui:2015kmb,Champion:2015pmj,Kulkarni:2015sxv,Popov:2018hkz}. Indeed, the corresponding radio and X-ray bursts can be naturally powered by flux-tube reconnection events in a strongly magnetized plasma of pulsars \cite{Lyutikov:2002kh}. The indirect evidence is provided by the detection of the first radio burst in the Milky Way from magnetar SGR 1935+2154 by CHIME/FRB \cite{CHIMEFRB:2020abu} and STARE2 \cite{Bochenek:2020zxn}, in temporal coincidence with high energy bursts observed by INTEGRAL \cite{Mereghetti:2020unm}, AGILE \cite{Tavani:2020ATe}, Konus-Wind \cite{Ridnaia:2020gcv}, and Insight-HXMT \cite{Insight-HXMTTeam:2020dmu}. The plasma parameters required to produce fast radio bursts are extreme, even when compared with brightest pulses from rotationally-powered pulsars \cite{Lyutikov:2019bcu}. Recently, it was proposed \cite{Philippov:2020jxu} that pulsar radio emission could be generated in non-stationary pair plasma discharges which are responsible for filling the pulsar magnetosphere with plasma.

The paper is organized as follows.  We will start with a general introduction to the physics of the magnetosphere in the next section. The role of the chiral anomaly in a relativistic electron-positron plasma will be reviewed in Sec.~\ref{sec:anomaly}. The estimates for the induced chiral charge and the role of chiral anomalous effects in magnetospheres of black holes and compact stars will be discussed in Sec.~\ref{sec:chiral-charge}. Physics implications of anomalous physics in magnetospheres and possible observable signatures are outlined in Sec.~\ref{sec:implications}. Finally, the summary of main results and conclusions are given in Sec.~\ref{sec:summary}. 

Throughout the paper we use units with $c=1$ and $\hbar=1$. 

\section{Fields and relativistic plasma in magnetospheres}
\label{sec:magnetosphere}

Black holes, compact stars, and their mergers are widely recognized as the ultimate natural laboratories of matter under extreme conditions. They contain superdense baryonic matter, can spin incredibly fast, and support some of the strongest magnetic fields in the present Universe. 

The theoretical possibility of the existence of neutron stars dates back to the 1930s \cite{Landau:1932,Baade:1934zex}, but the observational evidence came only with the discovery of radio pulsars by Antony Hewish, Jocelyn Bell, and collaborators \cite{Hewish:1968bj}. More than half a century later, the radio pulsars remain at the forefront of research, and many details of their physics are shrouded in mystery. Despite much progress, the detailed understanding of the structure of pulsar magnetospheres and the underlying nature of fast radio bursts remain enigmatic \cite{Melrose:2016kaf}. (For an instructive introduction and overview, see also Ref.~\cite{Beskin:2010iba}.) 

While acting on very different time and length scales, active galactic nuclei, which are powered by supermassive black holes, share some similarities with pulsars. They are also surrounded by magnetospheres and often have powerful polar jets extending over astronomical distances. The underlying mechanism of their activity is an active research area as well. As we will show below, however, the production of anomalous chiral charge is negligible in the black hole magnetospheres.

\begin{figure}
\resizebox{0.45\textwidth}{!}{%
  \includegraphics{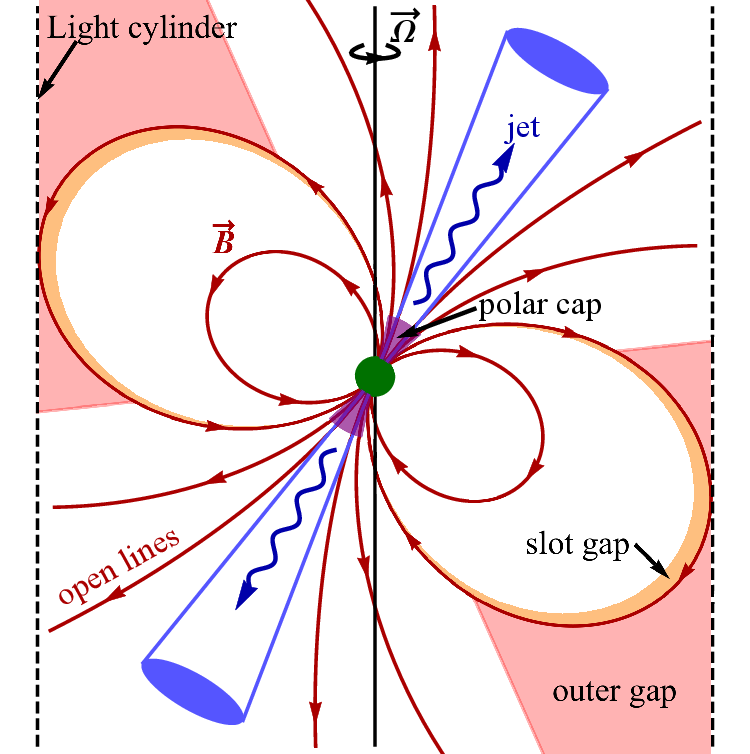}
}
\caption{A schematic illustration of the pulsar magnetosphere.}
\label{setup}  
\end{figure}

The consensus about the main features of the pulsar magnetospheres can be described as follows. The magnetic field of magnetosphere is that of an oblique rotator with a strongly magnetized charged plasma, see Fig.~\ref{setup} for a schematic illustration. The plasma is most likely populated with electrons and positrons. Within the light cylinder ($r<R_{LC}\equiv c/\Omega$, where $\Omega$ is the angular velocity), the plasma can be viewed approximately as rigidly co-rotating together with the star. Of course, the same assumption cannot be extended to the region outside the light cylinder ($r>R_{LC}$). 

The fact that the pulsar's magnetosphere is filled with a charged plasma rather than an empty vacuum is intimately connected with the presence of a strong magnetic field. It can be understood as follows. The unscreened electric field in a naive vacuum model would cause the emission of electrons and protons from the stellar surface. Because of the magnetic field, however, the charges will move along curved trajectories and radiate energetic photons. The latter, in turn, cause the secondary production of electron-positron pairs that accelerate and produce more photons. In the end, the magnetosphere fills out with a sufficient density of charged particles to screen out the electric field. The result is described approximately by the frozen-in magnetic field condition and the vanishing electric field in the co-rotating frame. The latter implies that $\bm{E} = - (\bm{\Omega}\times\bm{r})\times\bm{B}$, which is consistent with the vanishing parallel component of the electric field, $E_\parallel =0$. In this case, the charge density of the co-rotating plasma is given by the Goldreich-Julian relation \cite{Goldreich:1969sb}: $\rho_{GJ} \simeq - 2(\bm{\Omega}\cdot\bm{B})/\left[1-(r/R_{LC})^2\sin^2\theta\right]$, where $\theta$ is the angle measured from the rotation axis.

It is crucial, however, that all models of the rotating pulsar magnetospheres with the global force-free field configurations ($E_\parallel = 0$) are inconsistent with the Faraday law. (For a  concise review, see Ref.~\cite{Melrose:2016kaf}.) This can be seen from a charge or current starvation in certain regions of the model magnetospheres \cite{Sturrock:1971zc,Ruderman:1975ju,Arons:1983aa,Cheng:1986qt}. Therefore, the transient gap regions with $E_\parallel \neq 0$ must develop in realistic magnetospheres.

\subsection{Gap regions}
\label{sec:gap-regions}

The need for gaps in the conducting pulsar magnetosphere can be understood from a very general consideration. Over a sufficiently long time, the rotation of a pulsar will necessarily wind up the magnetic field flux tubes and build up stress in the field configuration. Eventually, the magnetic field lines should reconnect and some of the magnetic energy density should be released. Topologically, the reconnection of the magnetic field lines is characterized by the linking number. The latter cannot change unless a nonzero $\bm{E}\cdot\bm{B}$ occurs somewhere. 

Generically, the gaps in magnetospheres are intermittent as they will close after a sufficient density of the electron-positron pairs is produced. The gaps can appear near the polar regions of the stellar surface (the polar cap gap), near the light cylinder (the outer gap), or near the boundary between the open and closed magnetic field lines (the slot gap), see Fig.~\ref{setup}. We will concentrate primarily on the possible gaps near the polar regions where the magnetic field is the strongest. The corresponding gaps develop primarily because of the rapid outflow of charge from the magnetosphere along the open field lines \cite{Ruderman:1975ju}. The gap size $h$ grows at a speed close to the speed of light. Along with it, the electric potential difference across the gap grows like $h^2$ and, eventually, causes an avalanche of electron-positron pairs.

As we will discuss in Sec.~\ref{sec:anomaly}, the presence of parallel electric and magnetic fields can have profound implications for the dynamics of relativistic plasma in the pulsar magnetosphere. Because of the quantum chiral anomaly, they produce a transient out-of-equilibrium state with a nonzero chiral charge that, in turn, causes the chiral plasma instability and leads to the inverse cascade of magnetic helicity (see Sec.~\ref{sec:implications}). To quantify the corresponding processes, one needs to know the typical strengths of the electric fields inside the gap regions. Unlike the magnetic fields, which are relatively well known, the electric field inside the intermittent gaps is harder to estimate \cite{Ruderman:1975ju,Arons:1983aa,Cheng:1986qt}.

\subsection{Estimate of $E_{||}$}
\label{sec:electric-field}

According to Ref.~\cite{Ruderman:1975ju}, the parallel electric field in a gap is estimated to be $E_\parallel \simeq B h/R_{LC}$. The size of a polar gap is given by \cite{Ruderman:1975ju}
\begin{equation}
h\simeq 3.6~\mbox{m}\left(\frac{R}{10~\mbox{km}}\right)^{2/7}\left(\frac{\Omega}{1~\mbox{s}^{-1}}\right)^{-3/7}\left(\frac{B}{10^{14}~\mbox{G}}\right)^{-4/7},
\label{h-size}   % eq. (22) in Ref.~\cite{Ruderman:1975ju}  or eq.(53) in Ref.~\cite{Medin:2007vd}
\end{equation}
which is much smaller than the stellar radius \cite{Sturrock:1971zc,Ruderman:1975ju}. In the case of magnetars with superstrong magnetic fields, $B\gtrsim B_{c}\equiv m_e^2/e \approx  4.4 \times 10^{13}~\mbox{G}$, the above estimate should be corrected to account for the increase of the surface matter binding due to the magnetic field \cite{Medin:2007vd}. As a result, the estimate for $h$ may increase by a factor of $2$ to $3$ for the magnetars with magnetic fields that are $10$ to $100$ times stronger than $B_{c}$. The corresponding value of $E_\parallel$, which scales linearly with $h$, will also increase by the same factor. In this study, however, we will use the conservative estimate in Eq.~(\ref{h-size}).

By making use of Eq.~(\ref{h-size}), the magnitude of the parallel electric field in the gap is estimated as follows:
\begin{equation}
E_\parallel \approx 2.7\times 10^{-8}E_c
\left(\frac{R}{10~\mbox{km}}\right)^{2/7}\left(\frac{\Omega}{1~\mbox{s}^{-1}}\right)^{4/7} \left(\frac{B}{10^{14}~\mbox{G}}\right)^{3/7},
\label{eq:electric-field}
\end{equation}
where, by definition, the critical electric field is $E_c\equiv m_e^2/e \approx 1.3\times 10^{18}~\mbox{V/m}$.
One should note, however, that some models predict even larger values of the parallel electric field. According to Ref.~\cite{Petri:2016tqe}, for example, they could be as large as 
\begin{equation}
E_\parallel \simeq BR/R_{LC}\simeq 7.6 \times 10^{-5}E_c
\left(\frac{\Omega}{1~\mbox{s}^{-1}}\right)  \left(\frac{B}{10^{14}~\mbox{G}}\right) ,
\label{eq:electric-field-max}
\end{equation}
i.e., enhanced by a factor of $R/h$, which is about $10^{3}$ to $10^{4}$. In the most extreme case of young, short period magnetars, it was speculated that the electric field $E_{||}$ could approach up to 5$\%$ of the critical field $E_c$ before the gap closes by the copious production of the electron-positron pairs via the Schwinger process \cite{Lieu:2016hfw}. If such a scenario is realized, the value for $E_\parallel$ will be up to $6$ orders of magnitude larger than the estimate in Eq.~(\ref{eq:electric-field}).

\section{Chiral anomaly}
\label{sec:anomaly}

Before proceeding to the study of chiral dynamics in compact stars and black holes, let us discuss the general features of the chiral anomaly and its role in relativistic plasmas. Historically, the chiral anomaly showed up in particle physics in the theoretical description of the neutral pion decay rate. The anomaly also provides a key ingredient for the generation of an abnormally large mass of the $\eta^\prime$ meson.

Formally, the chiral anomaly implies that the classical chiral symmetry of a massless gauge theory is not the symmetry of the quantum action \cite{Adler:1969gk,Bell:1969ts}. Mathematically, this is described by the non-conservation of the chiral charge in the presence of parallel electric and magnetic fields:
\begin{equation}
\frac{\partial n_5}{\partial t} +\bm{\nabla}\cdot\bm{j}_5 = \frac{e^2}{2\pi^2} \mathbf{E}\cdot\mathbf{B}  ,
\label{anomaly}
\end{equation}
where $n_5$ is the chiral charge and $\bm{j}_5$ is the corresponding current density. (For our purposes, it is sufficient to consider an Abelian gauge theory, but the anomaly relation can be extended to a non-Abelian theory as well.)

Over the last decade or so, it came to light that, in addition to the well-known microscopic consequences in particle physics, the chiral anomaly can also lead to macroscopic phenomena in relativistic matter. (For reviews, see Refs.~\cite{Kharzeev:2013ffa,Liao:2014ava,Miransky:2015ava,Huang:2015oca,Kharzeev:2015znc}.)

Before discussing any physics implications of the anomalous continuity relation (\ref{anomaly}) in relativistic plasmas, it is instructive to recall that none of the known Dirac particles are truly massless. Thus, it is natural to ask what the role of a nonzero mass is. It is reasonable to expect that the mass should not play a big role in a relativistic plasma at a sufficiently high temperature ($T\gg m_e$) or  density ($\mu\gg m_e$, where $\mu$ is the chemical potential). Nevertheless, even a very small mass may trigger chirality flipping processes. In some cases, the corresponding rates may be significant. 

To understand better the role of chirality flipping, consider a uniform plasma in background electric and magnetic fields. Then the modified version of the anomalous continuity relation (\ref{anomaly}) takes the 
form
\begin{equation}
\frac{\partial n_5}{\partial t} = \frac{e^2}{2\pi^2} \mathbf{E}\cdot\mathbf{B} -\Gamma_m n_5 ,
\label{dn5-dt}
\end{equation}
where $\Gamma_m$ is the chiral flip rate due to a nonzero electron mass $m_e$. The assumption of a relativistic regime in the plasma is self-consistent when the temperature (or the chemical potential) is sufficiently large, i.e., $T\gg m_e$ (or, $\mu\gg m_e$, respectively). 

The anomalous chiral production rate, which is proportional to $\mathbf{E}\cdot\mathbf{B}$, is temperature independent. The chiral flip rate $\Gamma_m$, on the other hand, is temperature dependent \cite{Boyarsky:2020cyk,Boyarsky:2020ani}, i.e.,
\begin{equation}
\Gamma_m \simeq \alpha^2 \frac{m_e^2}{T}  ,
\label{flip-rate}
\end{equation}
where $\alpha= e^2/(4\pi)$ is the fine structure constant. Note that the flip rate is quadratic in $\alpha$ and, thus, relatively small. This estimate is justified when the plasma temperature is not too high, i.e., $ T\lesssim m_e/\sqrt{\alpha}\sim 10~\mbox{MeV}$. In the ultrarelativistic regime (i.e., $ T\gg m_e/\sqrt{\alpha}$), on the other hand, the rate becomes linear in $\alpha$ \cite{Boyarsky:2020cyk,Boyarsky:2020ani}. (In the case of cold relativistic matter, the flip rate $\Gamma_m$ is given by a similar expression, but the temperature $T$ should be replaced by the chemical potential $\mu$.)

Since the flip rate $\Gamma_m$ is small in the relativistic regime, the chiral charge of a quantum particle is a good approximate quantum number despite the nonzero electron mass. Furthermore, in a plasma made of such particles, the local chiral charge is also approximately conserved (up to the anomalous production). Like the local electric charge, therefore, it will also evolve slowly in a macroscopic system. This is the default regime of chiral matter. If the background electric and magnetic fields are uniform and evolving slowly, one finds from Eq.~(\ref{dn5-dt}) that the chiral charge density will tend to approach the following stationary value:
\begin{equation}
n_{5} = \frac{e^2}{2\pi^2\Gamma_m} \mathbf{E}\cdot\mathbf{B} .
\label{n5-stationary}
\end{equation}
In a state with this chiral charge density, the anomalous production of chirality by the background fields matches the loss of chirality via the chirality flipping processes. Importantly, the characteristic time scale $t^{\star}$ for reaching the quasi-stationary regime is very short compared to the dynamics of the macroscopic gap itself. Indeed, as is clear from Eq.~(\ref{dn5-dt}), it is determined by the inverse flip rate, $t^{\star}\simeq \Gamma_m^{-1}$, which is about $10^{-17}~\mbox{s}$ when $T\simeq 1~\mbox{MeV}$. In comparison,  the evolution time scale of the gap regions is much longer as it is set by their size, $t_h\simeq h$. According to the estimate in Eq.~(\ref{h-size}), it is about $ 10^{-8}~\mbox{s}$ when the magnetic field is $B\simeq 10^{14}~\mbox{G}$.

By recalling the standard interpretation of the anomalous chiral charge production in the background fields as a spectral flow, one may ask how such a process is possible at all when the electron mass is nonzero. Naively, there should be no room for a continuous spectral flow that turns the negative energy states (antiparticles) into positive energy states (particles) when $m_e\neq 0$. However, the direct calculations in Ref.~\cite{Copinger:2018ftr} show that the chiral charge production is nonzero although exponentially suppressed, i.e., $\partial n_5/\partial t \propto (\bm{E}\cdot\bm{B}) \exp(-\pi m_e^2/|eE|)$, provided the electric field is not too strong ($|eE|\ll m_e^2$). This shows that the anomalous spectral flow is effectively open in the same way as the Schwinger production of electron-positron pairs. 

The situation changes dramatically inside a hot relativistic plasma or in the presence of a flux of energetic photons. When the energy scale is comparable or exceeds the mass gap, the electron-positron pairs are produced constantly by photons or thermal excitations, and the chirality production is no longer suppressed exponentially. In this case, it is justifiable to use Eq.~(\ref{dn5-dt}) together with the flip rate in Eq.~(\ref{flip-rate}). 

Before concluding this section, it is interesting to note that the anomalous production of chiral charge by background electric and magnetic fields is not the only possible production mechanism. Because of the chiral nature of the electroweak theory, one may argue that a nonzero chiral charge density can be generated in a relativistic matter by electroweak processes. As speculated in Refs.~\cite{Charbonneau:2009ax,Ohnishi:2014uea,Kaminski:2014jda}, for example, a chiral imbalance could be produced in the core-collapse supernovae via the capture of left-handed electrons. However, a careful analysis shows that the resulting chiral charge density is minuscule \cite{Grabowska:2014efa} (see, however, Refs.~\cite{Dvornikov:2014uza,Sigl:2015xva}). It is due to the smallness of the electroweak rate compared to the flip rate given by $\Gamma_m \simeq \alpha^2 m_e^2/\mu_e$ in proto-neutron stars, where $\mu_e$ is the corresponding electron chemical potential. For the same reason, the weak processes will not play an important role in the magnetospheres of black holes and pulsars. 

\section{Chirality production in magnetosphere}
\label{sec:chiral-charge}

 Both supermassive black holes and pulsars can have strong polar jets powered by the magnetic energy of plasma in their magnetospheres. The energy release mechanism is likely to be connected with the intermittent gap regions, where magnetic flux tubes reconnect. As we discussed in Sec.~\ref{sec:gap-regions}, the value of $\bm{E}\cdot\bm{B}$ is nonzero in the gaps. Therefore, it is reasonable to suggest that, together with the rapid production of electron-positron plasma in the gaps, a substantial amount of chiral charge can be produced as well. Below we study the corresponding production rate and estimate the resulting values of the chiral chemical potential for both object types.

\subsection{Black holes}
\label{sec:black-holes}

Because of the no-hair theorem, black holes do not possess a magnetic field themselves. However, nonzero magnetic fields outside black holes can arise due to the accretion disks. Therefore, it is natural to expect that stellar-mass black holes should have much weaker magnetic fields than those in pulsars. However, if an exceptionally strong magnetic field is produced by a merger of a black hole and a strongly magnetized neutron star \cite{East:2021spd}, the estimate of the chiral charge production in their magnetosphere during the transition period can be obtained from the same analysis as in Sec.~\ref{sec:magnetars}. In this subsection, we consider only supermassive black holes in the active galactic nuclei whose accretion disks are of an astronomical size and could generate relativistic jets with very large Lorentz factors naively suggesting that a sufficiently large chiral charge can appear there.

To estimate the chirality production in a gap region of the magnetosphere near a supermassive black hole, we note that the corresponding magnetic fields are of the order of $B\simeq 10^2~\mbox{G}$ \cite{Eatough:2013nva,EventHorizonTelescope:2021srq}. As a specific example, one can use the parameters of the black hole at the center of the elliptic galaxy M87 in the constellation Virgo. Its shadow was observed recently by the Event Horizon Telescope \cite{EventHorizonTelescope:2019pgp}. The mass of the black hole is measured to be $6.5\times 10^9\,M_{\odot}$ and its Schwarzschild radius is $R_s=120~\mbox{au}$. We will use the latter to set the characteristic time scale for the plasma processes, $t_0\sim R_s\simeq 1.8\times10^{13}~\mbox{m}$.

It is instructive to estimate first the chiral charge density due to the chiral anomaly in the absence of chirality flipping. We find
\begin{equation}
n_5^{\rm naive} \simeq \frac{e^2}{2\pi^2} \mathbf{E}\cdot\mathbf{B}  t  
\simeq 1.6\times 10^{-3}~\mbox{MeV}^3 \left(\frac{t}{t_0}\right) \left(\frac{B}{10^{2}~\mbox{G}}\right)^2 ,
\end{equation}
where we used a naive optimistic estimate $E\simeq 10^{-3} B$ \cite{Beskin:1992aaa,Ptitsyna:2015nta}. The above equation suggests that $n_5$ and, therefore, $\mu_5$ can be very large.

The inclusion of the chiral flip changes the situation dramatically however. Indeed, by making use of Eqs.~(\ref{flip-rate}) and (\ref{n5-stationary}), one finds that the chiral charge density is bound from above by 
\begin{eqnarray}
n_5 &\simeq&  \frac{e^2}{2\pi^2 \Gamma_m} \mathbf{E}\cdot\mathbf{B} \nonumber\\
&\simeq& 1.3\times 10^{-24}~\mbox{MeV}^{3}\left(\frac{T}{1~\mbox{MeV}}\right)\left(\frac{B}{10^{2}~\mbox{G}}\right)^2 ,
\label{eq:n5-BH}
\end{eqnarray}
where we again assumed that $E\simeq 10^{-3} B$. It should be noted that the (co-moving)  plasma temperature in the magnetospheres of black holes might be considerably smaller than $1~\mbox{MeV}$. Then, the estimate for $n_5$ will be even smaller. Additionally, the assumption of the relativistic regime may not be fully justified. 

By using the definition for the chiral charge density in terms of the chemical potential $\mu_5$, i.e.,
\begin{equation}
n_5 = \frac{\mu_5\left(\pi^2 T^2+ 3 \mu_5^2\right)}{3 \pi^2}
\end{equation}
and taking into account that  $\mu_5\ll T$, we obtain the following value of the chiral chemical potential:
\begin{equation}
\mu_5 \simeq \frac{3n_5}{T^2}
\simeq  3.8\times 10^{-24}~\mbox{MeV}~\left(\frac{T}{1~\mbox{MeV}}\right)^{-1} \left(\frac{B}{10^{2}~\mbox{G}}\right)^2.
\label{eq:mu5-BH}
\end{equation}
By any standard, this is negligible. Since magnetic fields in pulsars and especially in magnetars are much stronger than in black holes, there is a hope that the chiral asymmetry in their magnetospheres is much larger. We investigate this possibility in the next subsection.

\subsection{Magnetars}
\label{sec:magnetars}

From the viewpoint of chirality production, the gap regions of pulsar magnetospheres are much more promising. First of all, their much stronger magnetic fields can drastically increase the anomalous production of chiral charge. Additionally, the presence of a superstrong magnetic field helps to sustain a truly relativistic regime in the plasma \cite{Arendt:2002ay}. 

By making use of the estimates for the parallel electric field (\ref{eq:electric-field}) and the chiral flip rate (\ref{flip-rate}), we derive the following estimate for the chiral charge density:
\begin{eqnarray}
n_5 & \simeq & \frac{e^2E_\parallel B}{2\pi^2 \Gamma_m} 
\simeq 1.5\times 10^{-5}~\mbox{MeV}^3 \left(\frac{T}{1~\mbox{MeV}}\right)\nonumber\\
& \times & 
\left(\frac{R}{10~\mbox{km}}\right)^{2/7}\left(\frac{\Omega}{1~\mbox{s}^{-1}} \right)^{4/7}
\left(\frac{B}{10^{14}~\mbox{G}}\right)^{10/7}.
\label{eq:n5-pulsars}
\end{eqnarray}
The corresponding value of the chiral chemical potential is given by
\begin{eqnarray}
\mu_5 & \simeq & \frac{3n_5}{T^2} \simeq  4.6\times 10^{-5}~\mbox{MeV}~
\left(\frac{T}{1~\mbox{MeV}}\right)^{-1}\nonumber\\
& \times & 
\left(\frac{R}{10~\mbox{km}}\right)^{2/7}\left(\frac{\Omega} {1~\mbox{s}^{-1}}\right)^{4/7}
\left(\frac{B}{10^{14}~\mbox{G}}\right)^{10/7}.
\label{eq:mu5-pulsars}
\end{eqnarray}
By comparing it with the estimate for black holes in Eqs.~(\ref{eq:n5-BH}) and (\ref{eq:mu5-BH}), we see that a much larger chiral asymmetry is produced in the gap regions of magnetars. Mostly, this is the result of a strong magnetic field. For the production of chirality, it is also helpful that the electric field in the gaps grows with the magnetic field, see Eq.~(\ref{eq:electric-field}). 

We summarize the model parameters and the estimated values of chiral characteristics induced in the gap regions of pulsar magnetospheres in Table~\ref{tab:chiral}. The four columns of estimates correspond to four different values of the surface magnetic fields. As expected, the amount of chiral asymmetry grows with the field strength. As we see, the chiral chemical potential can reach values up to about $\mu_5\simeq 10^{-3}~\mbox{MeV}$ when $B\simeq 10^{15}~\mbox{G}$. As we will discuss in Sec.~\ref{sec:implications}, it might be sufficiently large to have observable implications in magnetars. 

\begin{table}
\caption{Estimated values of the model parameters and the chiral characteristics in the gap regions of the pulsar magnetospheres.}
\label{tab:chiral}
  \begin{tabular}{c|cccc}
\hline\noalign{\smallskip}
$B$  & $10^{12}~\mbox{G}$ & $10^{13}~\mbox{G}$  & $10^{14}~\mbox{G}$   & $10^{15}~\mbox{G}$   \\ 
\noalign{\smallskip}\hline\noalign{\smallskip}
                                 $h$                     &   $50~\mbox{m}$    & $13.4~\mbox{m}$    & $3.6~\mbox{m}$       & $0.97~\mbox{m}$ \\ 
      $\frac{E_\parallel}{E_c}$               & $3.8\times 10^{-9}$ & $1.0\times 10^{-8} $ & $2.7\times 10^{-8}$  & $7.3\times 10^{-8}$  \\ 
 $\frac{\bm{E}\cdot\bm{B}}{E_cB_c}$ & $8.6\times 10^{-11}$ & $2.3\times 10^{-9} $ & $6.2\times 10^{-8}$  & $1.7\times 10^{-6}$  \\
             $\frac{n_5}{m_e^3}$              & $1.6\times 10^{-7}$ & $4.3\times 10^{-6} $ & $1.1\times 10^{-4}$  & $3.1\times 10^{-3}$  \\             
             $\frac{\mu_5}{m_e}$              & $1.2\times 10^{-7}$ & $3.4\times 10^{-6} $ & $9.0\times 10^{-5}$  & $2.4\times 10^{-3}$  \\
            $\frac{k_\star}{m_e}$              & $5.8\times 10^{-10}$ & $1.6\times 10^{-8} $ & $4.2\times 10^{-7}$  & $1.1\times 10^{-5}$  \\
\noalign{\smallskip}\hline
\end{tabular}
\end{table}
 
 We should emphasize that all estimates above were obtained by utilizing the field-independent chiral flip rate in Eq.~(\ref{flip-rate}). It is likely, however, that the corresponding rate is modified in the presence of fields comparable to or stronger than the Schwinger critical value $B_c$. In such a regime, the electron and positron states are described by the Landau levels rather than plane waves. Moreover, by recalling that the lowest Landau level is spin-polarized, one may speculate that the chiral flip rate $\Gamma_m$ can be strongly suppressed compared to the naive estimate in Eq.~(\ref{flip-rate}). Without doing the actual calculation, this is indeed reasonable since the electron kinematics in the lowest Landau level is dimensionally reduced and, thus, highly constrained. If a strong magnetic field does suppress $\Gamma_m$, the chiral chemical potential in the magnetar magnetospheres can be greater than the estimate in Eq.~(\ref{eq:mu5-pulsars}).

\section{Physics implications of chiral charge}
\label{sec:implications}

Let us now discuss possible physics consequences of the chiral charge produced in the gaps of pulsar magnetospheres. Similar to the gap regions themselves, the associated chiral phenomena are intermittent and transient. Below we will consider a single gap region with a nonzero chiral chemical potential $\mu_5$, induced by the chiral anomaly. 

A hot relativistic plasma with a nonzero chiral charge density develops the chiral plasma instability, which in turn leads to the inverse cascade of magnetic helicity \cite{Tuchin:2014iua,Joyce:1997uy,Boyarsky:2011uy,Tashiro:2012mf,Manuel:2015zpa,Hirono:2015rla,Sigl:2015xva,Akamatsu:2013pjd}. The underlying mechanism of the instability is the spontaneous generation of helical electromagnetic modes with long wavelengths. 

To see how the instability develops, let us note that the electric current in a conducting chiral plasma is given by 
\begin{equation}
\bm{j} = \frac{2\alpha}{\pi} \mu_5 \mathbf{B}+\sigma \mathbf{E},
\label{current}
\end{equation}
where the first term represents a non-dissipative contribution due to the chiral magnetic effect (for a review, see Ref.~\cite{Kharzeev:2013ffa}). This current is saturated by the states in the lowest Landau level and, as a result, it is topologically protected. The second term in Eq.~(\ref{current}) is the usual (dissipative) Ohm's current. The conductivity of the plasma is denoted by $\sigma$. For the relativistic electron-positron plasma, it is given by $\sigma\simeq 6 T/e^2$ at the leading order in coupling \cite{Arnold:2003zc}.

By substituting Eq.~(\ref{current}) into the Ampere law\footnote{In a more refined approximation, one needs to account for the medium effects in the Maxwell equations by using a field-modified polarization tensor. While the technical consideration would get considerably more complicated, the general qualitative results should not change dramatically.}
\begin{equation}
\bm{\nabla}\times\mathbf{B} = \bm{j}+ \frac{\partial \mathbf{E}}{\partial t} , 
\label{Maxwell-eq2}
\end{equation}
and solving for the electric field, one arrives at the following results:
\begin{equation}
\mathbf{E} =\frac{1}{\sigma}\left( \bm{\nabla}\times\mathbf{B}  - k_\star\mathbf{B} - \frac{\partial \mathbf{E}}{\partial t} \right),
\label{EBj}
\end{equation}
where we introduced the shorthand notation $k_\star\equiv 2\alpha\mu_5/\pi$. As we will see below, the value of $k_\star$ determines the characteristic scale for wave vectors of the chiral plasma instability.

Note that, in this section, $\mathbf{E}$ and $\mathbf{B}$ include both background and dynamical fields.
By applying the curl operator on both sides of Eq.~(\ref{EBj}) and using the Faraday law, 
\begin{equation}
\bm{\nabla}\times\mathbf{E} = - \frac{\partial \mathbf{B}}{\partial t},
\label{Maxwell-eq3}
\end{equation} 
we obtain the final master equation for the evolution of magnetic field in a chirally imbalanced plasma 
\begin{equation}
\frac{\partial \mathbf{B}}{\partial t} 
=- \frac{1}{\sigma}\left( \bm{\nabla}\times(\bm{\nabla}\times\mathbf{B} ) - k_\star   \bm{\nabla}\times\mathbf{B} + \frac{\partial^2 \mathbf{B}}{\partial t^2}\right).
\label{B-time-dependence}
\end{equation} 
It is instructive to mention that the background magnetic field drops out from this equation and, therefore, has no direct effect on the development of the chiral plasma instability. Indirectly, of course, the value of $k_\star$ is determined by the initial background fields via $\bm{E}\cdot\bm{B}$ when the gap was open. To a good approximation, the transient background electric field $\bm{E}_\parallel$ is largely gone by the time when the instability develops. Thus, we set it to zero below.

To reveal the instability and identify the corresponding electromagnetic modes, it is convenient to decompose the field into helical eigenstates, i.e., $\mathbf{B} =\sum \alpha_{\lambda,k} \mathbf{B}_{\lambda,k}$. The latter are defined as the eigenstates of the curl operator, i.e., 
\begin{equation}
\bm{\nabla}\times\mathbf{B}_{\lambda,k} = \lambda k \mathbf{B}_{\lambda,k},
\end{equation} 
where $\lambda=\pm 1$ represents the helicities of the modes. A simple example of a helical mode is given by the following circularly polarized plane wave:
\begin{equation}
\mathbf{B}_{\lambda,k} = B_0 \left(\hat{\bm{x}}+ i  \lambda \hat{\bm{y}}\right)e^{-i\omega t +i k z}.
\end{equation} 
The conclusions do not depend on the specific choice of helical eigenstates, however. One can also use, for instance, the Chandrasekhar-Kendall states \cite{Chandrasekhar:1957}.

Since the master equation (\ref{B-time-dependence}) is linear, it can be analyzed mode by mode. Then, the field evolution of a single mode with helicity $\lambda$ reads
\begin{equation}
\frac{d \mathbf{B}_{\lambda,k}}{dt} = \frac{1}{\sigma}\left(\lambda k_\star   k -k^2 - \frac{\partial^2 }{\partial t^2} \right)\mathbf{B}_{\lambda,k}.
\label{time-B-eq}
\end{equation} 
Strictly speaking, such a treatment is an approximation where the backreaction of the dynamical fields on the chiral asymmetry is neglected. In a more systematic approach, of course, the production and growth of helical modes will be gradually depleting the chiral charge and reducing $\mu_5$  \cite{Boyarsky:2011uy,Manuel:2015zpa,Hirono:2015rla,Sigl:2015xva}. The corresponding evolution equation for $\mu_5$ in a given gap of volume $V$ follows from Eq.~(\ref{dn5-dt}) after including the dynamical electromagnetic fields, i.e.,
\begin{equation}
\frac{d\mu_5}{dt} \simeq -\frac{3\alpha}{\pi T^2} \sum_{\lambda=\pm 1}\frac{\lambda}{V} \int \frac{d^3\bm{k}}{(2\pi)^3 k } \frac{\partial |\mathbf{B}_{\lambda,k}|^2}{\partial t} -\Gamma_m \mu_5,
\label{dmu5-dt-helical}
\end{equation} 
where we used the relation between the chiral charge and the chiral chemical potential in Eq.~(\ref{eq:mu5-BH}). 

By assuming that $\mathbf{B}_{\lambda,k} \propto B_0  e^{-i \omega t}$ in Eq.~(\ref{time-B-eq}), we arrive at the following solutions for the frequency:
\begin{eqnarray}
\omega_{1,2} &=&- \frac{i}{2} \left(\sigma \pm \sqrt{\sigma^2 +4k(\lambda k_\star-k)}\right)
\nonumber\\
&\simeq & \left\{
\begin{array}{l}
-i \left(\sigma+\frac{k(\lambda k_\star-k)}{\sigma} \right), \\
i \frac{k(\lambda k_\star-k)}{\sigma},
\end{array}
\right.
\label{two-modes}
\end{eqnarray}
where the last approximation corresponds to the limit of high conductivity. [While the approximation of high conductivity is not necessary, it helps to identify the unstable mode more easily.]

As we see from Eq.~(\ref{two-modes}), one of the modes is $\mathbf{B}_{k,1} \propto B_0 e^{-\sigma t}$, which is damped by high conductivity. This is the usual charge diffusion mode in the plasma. The other mode, i.e., $\mathbf{B}_{k,2} \propto B_0 e^{t k(\lambda k_\star-k)/\sigma}$, is very different. For one of the helicities, determined by $\lambda =\mbox{sign}(\mu_5)$, the amplitude of the second mode grows exponentially, provided its wave vector is smaller than $|k_\star|$. Of course, this is an indication of instability in the plasma. As the amplitude of the helical mode grows, it gains energy. This energy comes from the chiral electron plasma. Recall that the plasma with a nonzero $\mu_5$ is out of equilibrium. Its free energy density is greater than the equilibrium value by $\Delta \epsilon \simeq  \mu_5^2 T^2$. The corresponding energy surplus was pumped into the plasma by the transient electric field in the gap. After the gap closes, the extra energy feeds the instability. The resulting maximum field strength of helical fields $|\bm{B}_{\lambda,k}| \lesssim \mu_5 T$ is small compared to the background magnetic field.

In view of Eq.~(\ref{dmu5-dt-helical}), the value of $\mu_5$ decreases that, in turn, attenuates the chiral plasma instability \cite{Boyarsky:2011uy,Manuel:2015zpa,Hirono:2015rla,Sigl:2015xva}. Indeed, when the limit $\mu_5\to 0$ is approached, the second mode turns into a diffusive mode with $\mathbf{B}_{k,2} \propto B_0 e^{-t k^2/\sigma}$ for small wave vectors ($k\ll \sigma/2$). As we see, the rate of its decay is proportional to $k^2$ and inversely proportional to the conductivity. 

From the dispersion relation of the unstable mode in Eq.~(\ref{two-modes}), it is easy to determine that the fastest growing mode has the wave vector 
\begin{eqnarray}
k_{*} &=& \frac{|k_\star|}{2} = \frac{\alpha}{\pi} |\mu_5|   \simeq  1.1\times 10^{-7}~\mbox{MeV}~
\left(\frac{T}{1~\mbox{MeV}}\right)^{-1}\nonumber\\
& \times & 
\left(\frac{R}{10~\mbox{km}}\right)^{2/7}\left(\frac{\Omega}{1~\mbox{s}^{-1}} \right)^{4/7}
\left(\frac{B}{10^{14}~\mbox{G}}\right)^{10/7}.
\label{eq:kstar-pulsars}
\end{eqnarray}
A more careful analysis of chiral dynamics in a similar type of primordial plasma with nonzero $\mu_5$~\cite{Rogachevskii:2017uyc,Brandenburg:2017rcb,Schober:2017cdw} suggests that the initial chiral plasma instability may trigger the chiral magnetically driven turbulence and lead to a large-scale dynamo instability. In the end, the saturation of magnetic helicity and the growth of magnetic modes are controlled by the total chiral charge. In the case of the pulsar magnetosphere, however, the dynamics could be quenched sooner if the relevant correlation length reaches the size of the gap region. In either case, it is expected that the unstable plasma in the gap region will undergo the inverse magnetic cascade and radiate the energy in the form of helical modes in a wide range of frequencies, i.e., $0\lesssim \omega  \lesssim k_{\star}$. By making use of the estimate in Eq.~(\ref{eq:kstar-pulsars}), we see that the corresponding window of frequencies spans all radio waves and may reach into the near-infrared range. 

It is instructive to mention that the onset of the chiral plasma instability is a relatively slow process. Indeed, the corresponding characteristic time scale $t_{\rm inst}$ is determined by the inverse imaginary frequency of the unstable mode in Eq.~(\ref{two-modes}). For the wave vector in Eq.~(\ref{eq:kstar-pulsars}), we estimate $t_{\rm inst} \simeq 2 \sigma/k_{*}^2 \simeq 7\times 10^{-6}~\mbox{s}$ when $T\simeq 1~\mbox{MeV}$. Here we utilized the leading order result for the electrical conductivity $\sigma\simeq 6T/e^2$ obtained in Ref.~\cite{Arnold:2003zc}. As one might expect, time $t_{\rm inst}$ is large compared to the time needed for the anomaly to generate a nonzero $\mu_5$.

By noting that the chiral imbalance of the plasma feeds the instability, we can estimate the total energy available for the corresponding helical modes. In a region of size $h$, the energy associated with the chiral charge imbalance is given by $\Delta {\cal E} \simeq  \mu_5^2 T^2 h^3$, where we took into account that $\mu_5\ll T$. Thus, by making use of the result for the gap size in Eq.~(\ref{h-size}) and the chiral chemical potential in Eq.~(\ref{eq:mu5-pulsars}), we derive the following estimate:
\begin{eqnarray}
\Delta {\cal E} &\simeq&  2.1\times 10^{25}~\mbox{erg}~
\left(\frac{T}{1~\mbox{MeV}}\right) \left(\frac{R}{10~\mbox{km}}\right)^{6/7} \nonumber\\
& \times & 
\left(\frac{\Omega}{1~\mbox{s}^{-1}} \right)^{-9/7}
\left(\frac{B}{10^{14}~\mbox{G}}\right)^{2/7}.
\label{eq:energy-chiral-pulsars}
\end{eqnarray}
As we see, there is a substantial amount of energy that goes into the helical electromagnetic modes. If taken at face value, this much energy  certainly can affect observable features of pulsars. However, one should accept the estimate with great caution because only a fraction of the available energy can probably go into electromagnetic radiation. To study the problem in detail, one would need to perform carefully magnetohydrodynamic simulations, which are beyond the scope of this paper.

\section{Summary}
\label{sec:summary}

In this paper, we studied the possibility of  anomalous production of chiral charge in the gap regions of magnetospheres of pulsars and supermassive black holes. While naive estimates may suggest that a substantial amount of chiral charge is produced, the actual values are strongly limited by the chiral flip rate. 

In the case of supermassive black holes, it results in a very small chiral asymmetry, which is quantified by the estimates of $n_5$ and $\mu_5$ in Eqs.~(\ref{eq:n5-BH}) and (\ref{eq:mu5-BH}), respectively. To a large degree, the smallness of the induced chiral charge is explained by the relatively weak magnetic fields near the black holes. We do not expect that such a little chiral asymmetry can produce any observable effects.

The situation is qualitatively different in the case of pulsars and, especially, magnetars, which are characterized by very strong surface magnetic fields. In the gap regions of their magnetospheres, the anomalous chiral charge production is significant. In the case of magnetars, in particular, the estimated values of the chiral chemical potential can be of the order of $10^{-5}~\mbox{MeV}$ to $10^{-3}~\mbox{MeV}$, see Eq.~(\ref{eq:mu5-pulsars}). While this does not seem to be large compared to the electron rest-mass energy, it can have important implications for the physics of magnetars. 

We argue that the chiral charge produced in the gap regions of a pulsar magnetosphere is sufficient to trigger spontaneous generation and emission of helical electromagnetic modes in a wide window of frequencies, limited from above by the wave vector $k_\star \simeq \alpha \mu_5$. The corresponding range includes radio and near-infrared waves.

Considering the nature of the chiral plasma instability, we can speculate that the associated electromagnetic emission should be circularly polarized. Its handedness is set by the sign of the chiral charge and, therefore, correlated with the magnetic field direction. However, the situation may get more complicated because the unstable modes in the gaps are likely to be partially modified by the plasma dynamics in the bulk of the magnetosphere before the electromagnetic radiation is emitted into space.

In this exploratory study, we did not perform a detailed analysis of the complicated plasma dynamics in the gap regions. However, our qualitative considerations suggest that anomalous chiral production can play an important role in the evolution of plasma instabilities and the generation of radio emission from pulsars. The latter may affect the physical processes in the pulsar jets and modify observable features of the fast radio bursts.

While our preliminary study indicates that the anomalous chiral charge production can indeed modify the physics properties of the pulsar magnetospheres, a more detailed self-consistent analysis of all stages of plasma dynamics in the gaps should be performed in the future. Firstly, one should perform a careful simulation of the chiral charge generation during the electron-positron pair production in the gaps at the initial stages of the evolution. Secondly, a quantitative study of the magnetodynamics evolution of the chiral plasma instability has to be done for the specific conditions in the gap regions of the pulsar magnetospheres. 
 
From the theoretical viewpoint, one of the limitations of the current study is a rough estimate of the chiral flip rate $\Gamma_m$. We used the result obtained in the limit of the vanishing magnetic field. As we argued in the main text, however, such an approximation may overestimate the flip rate. Therefore, to improve the predictions for the chiral charge density, induced by the chiral anomaly in a strongly magnetized plasma, a rigorous study of the chiral flip rate at $B\neq 0$ will be required.

\section*{Acknowledgements}
The work of E.V.G. is supported in part by the Program of Fundamental Research of the Physics and Astronomy Division of the NAS of Ukraine. 
The work of I.A.S. was supported by the U.S. National Science Foundation under Grant No.~PHY-1713950.


\begin{thebibliography}{}

\bibitem{Kharzeev:2007jp}
D.~E.~Kharzeev, L.~D.~McLerran, and H.~J.~Warringa,
%``The Effects of topological charge change in heavy ion collisions: 'Event by event P and CP violation',''
Nucl. Phys. A \textbf{803}, 227 (2008),
% doi:10.1016/j.nuclphysa.2008.02.298
arXiv:0711.0950 [hep-ph].

\bibitem{Kharzeev:2010gr}
D.~E.~Kharzeev and D.~T.~Son,
%``Testing the chiral magnetic and chiral vortical effects in heavy ion collisions,''
Phys. Rev. Lett. \textbf{106}, 062301 (2011),
% doi:10.1103/PhysRevLett.106.062301
arXiv:1010.0038 [hep-ph].

\bibitem{Gorbar:2011ya}
E.~V.~Gorbar, V.~A.~Miransky, and I.~A.~Shovkovy,
%``Normal ground state of dense relativistic matter in a magnetic field,''
Phys. Rev. D \textbf{83}, 085003 (2011),
% doi:10.1103/PhysRevD.83.085003
arXiv:1101.4954 [hep-ph].

\bibitem{Burnier:2011bf}
Y.~Burnier, D.~E.~Kharzeev, J.~Liao, and H.~U.~Yee,
%``Chiral magnetic wave at finite baryon density and the electric quadrupole moment of quark-gluon plasma in heavy ion collisions,''
Phys. Rev. Lett. \textbf{107}, 052303 (2011),
% doi:10.1103/PhysRevLett.107.052303
arXiv:1103.1307 [hep-ph].

\bibitem{Bzdak:2012ia}
A.~Bzdak, V.~Koch, and J.~Liao,
%``Charge-Dependent Correlations in Relativistic Heavy Ion Collisions and the Chiral Magnetic Effect,''
Lect. Notes Phys. \textbf{871}, 503 (2013),
% doi:10.1007/978-3-642-37305-3\_19
arXiv:1207.7327 [nucl-th].

\bibitem{Kharzeev:2013ffa}
D.~E.~Kharzeev,
%``The Chiral Magnetic Effect and Anomaly-Induced Transport,''
Prog. Part. Nucl. Phys. \textbf{75}, 133 (2014),
% doi:10.1016/j.ppnp.2014.01.002
arXiv:1312.3348 [hep-ph].

\bibitem{Tuchin:2014iua}
K.~Tuchin,
%``Electromagnetic field and the chiral magnetic effect in the quark-gluon plasma,''
Phys. Rev. C \textbf{91}, 064902 (2015),
%doi:10.1103/PhysRevC.91.064902
arXiv:1411.1363 [hep-ph].

\bibitem{Liao:2014ava}
J.~Liao,
%``Anomalous transport effects and possible environmental symmetry \textquoteleft{}violation\textquoteright{} in heavy-ion collisions,''
Pramana \textbf{84}, 901 (2015),
% doi:10.1007/s12043-015-0984-x
arXiv:1401.2500 [hep-ph].

\bibitem{Miransky:2015ava}
V.~A.~Miransky and I.~A.~Shovkovy,
%``Quantum field theory in a magnetic field: From quantum chromodynamics to graphene and Dirac semimetals,''
Phys. Rept. \textbf{576}, 1 (2015),
% doi:10.1016/j.physrep.2015.02.003
arXiv:1503.00732 [hep-ph].

\bibitem{Huang:2015oca}
X.~G.~Huang,
%``Electromagnetic fields and anomalous transports in heavy-ion collisions --- A pedagogical review,''
Rept. Prog. Phys. \textbf{79}, 076302 (2016),
% doi:10.1088/0034-4885/79/7/076302
arXiv:1509.04073 [nucl-th].

\bibitem{Kharzeev:2015znc}
D.~E.~Kharzeev, J.~Liao, S.~A.~Voloshin, and G.~Wang,
%``Chiral magnetic and vortical effects in high-energy nuclear collisions\textemdash{}A status report,''
Prog. Part. Nucl. Phys. \textbf{88}, 1 (2016),
% doi:10.1016/j.ppnp.2016.01.001
arXiv:1511.04050 [hep-ph].

\bibitem{Joyce:1997uy}
M.~Joyce and M.~E.~Shaposhnikov,
%``Primordial magnetic fields, right-handed electrons, and the Abelian anomaly,''
Phys. Rev. Lett. \textbf{79}, 1193 (1997),
% doi:10.1103/PhysRevLett.79.1193
arXiv:astro-ph/9703005 [astro-ph].

\bibitem{Boyarsky:2011uy}
A.~Boyarsky, J.~Frohlich, and O.~Ruchayskiy,
%``Self-consistent evolution of magnetic fields and chiral asymmetry in the early Universe,''
Phys. Rev. Lett. \textbf{108}, 031301 (2012),
% doi:10.1103/PhysRevLett.108.031301
arXiv:1109.3350 [astro-ph.CO].

\bibitem{Boyarsky:2012ex}
A.~Boyarsky, O.~Ruchayskiy, and M.~Shaposhnikov,
%``Long-range magnetic fields in the ground state of the Standard Model plasma,''
Phys. Rev. Lett. \textbf{109}, 111602 (2012),
% doi:10.1103/PhysRevLett.109.111602
arXiv:1204.3604 [hep-ph].

\bibitem{Tashiro:2012mf}
H.~Tashiro, T.~Vachaspati, and A.~Vilenkin,
%``Chiral Effects and Cosmic Magnetic Fields,''
Phys. Rev. D \textbf{86}, 105033 (2012),
% doi:10.1103/PhysRevD.86.105033
arXiv:1206.5549 [astro-ph.CO].

\bibitem{Tashiro:2013bxa}
H.~Tashiro and T.~Vachaspati,
%``Cosmological magnetic field correlators from blazar induced cascade,''
Phys. Rev. D \textbf{87}, 123527 (2013),
% doi:10.1103/PhysRevD.87.123527
arXiv:1305.0181 [astro-ph.CO].

\bibitem{Tashiro:2013ita}
H.~Tashiro, W.~Chen, F.~Ferrer, and T.~Vachaspati,
%``Search for CP Violating Signature of Intergalactic Magnetic Helicity in the Gamma Ray Sky,''
Mon. Not. Roy. Astron. Soc. \textbf{445}, L41 (2014),
% doi:10.1093/mnrasl/slu134
arXiv:1310.4826 [astro-ph.CO].

\bibitem{Manuel:2015zpa}
C.~Manuel and J.~M.~Torres-Rincon,
%``Dynamical evolution of the chiral magnetic effect: Applications to the quark-gluon plasma,''
Phys. Rev. D \textbf{92}, 074018 (2015),
% doi:10.1103/PhysRevD.92.074018
arXiv:1501.07608 [hep-ph].

\bibitem{Hirono:2015rla}
Y.~Hirono, D.~E.~Kharzeev, and Y.~Yin,
%``Self-similar inverse cascade of magnetic helicity driven by the chiral anomaly,''
Phys. Rev. D \textbf{92}, 125031 (2015),
% doi:10.1103/PhysRevD.92.125031
arXiv:1509.07790 [hep-th].

\bibitem{Armitage:2017cjs}
N.~P.~Armitage, E.~J.~Mele, and A.~Vishwanath,
%``Weyl and Dirac Semimetals in Three Dimensional Solids,''
Rev. Mod. Phys. \textbf{90}, 015001 (2018),
%doi:10.1103/RevModPhys.90.015001
arXiv:1705.01111 [cond-mat.str-el].

\bibitem{Gorbar:2021ebc}
E.~V.~Gorbar, V.~A.~Miransky, I.~A.~Shovkovy, and P.~O.~Sukhachov,
{\it Electronic Properties of Dirac and Weyl Semimetals},
(World Scientific, Singapore, 2021).
%doi:10.1142/11475

\bibitem{Charbonneau:2009ax}
J.~Charbonneau and A.~Zhitnitsky,
%``Topological Currents in Neutron Stars: Kicks, Precession, Toroidal Fields, and Magnetic Helicity,''
JCAP \textbf{08}, 010 (2010),
% doi:10.1088/1475-7516/2010/08/010
arXiv:0903.4450 [astro-ph.HE].

\bibitem{Ohnishi:2014uea}
A.~Ohnishi and N.~Yamamoto,
%``Magnetars and the Chiral Plasma Instabilities,''
arXiv:1402.4760 [astro-ph.HE].

\bibitem{Kaminski:2014jda}
M.~Kaminski, C.~F.~Uhlemann, M.~Bleicher, and J.~Schaffner-Bielich,
%``Anomalous hydrodynamics kicks neutron stars,''
Phys. Lett. B \textbf{760}, 170 (2016),
%doi:10.1016/j.physletb.2016.06.054
arXiv:1410.3833 [nucl-th].

\bibitem{Dvornikov:2014uza}
M.~Dvornikov and V.~B.~Semikoz,
%``Magnetic field instability in a neutron star driven by the electroweak electron-nucleon interaction versus the chiral magnetic effect,''
Phys. Rev. D \textbf{91}, 061301(R) (2015),
% doi:10.1103/PhysRevD.91.061301
arXiv:1410.6676 [astro-ph.HE].

\bibitem{Sigl:2015xva}
G.~Sigl and N.~Leite,
%``Chiral Magnetic Effect in Protoneutron Stars and Magnetic Field Spectral Evolution,''
JCAP \textbf{01}, 025 (2016),
% doi:10.1088/1475-7516/2016/01/025
arXiv:1507.04983 [astro-ph.HE].

\bibitem{Dvornikov:2015iua}
M.~Dvornikov,
%``Relaxation of the chiral imbalance and the generation of magnetic fields in magnetars,''
J. Exp. Theor. Phys. \textbf{123}, 967 (2016),
% doi:10.7868/S0044451016120000
arXiv:1510.06228 [hep-ph].

\bibitem{Yamamoto:2015gzz}
N.~Yamamoto,
%``Chiral transport of neutrinos in supernovae: Neutrino-induced fluid helicity and helical plasma instability,''
Phys. Rev. D \textbf{93}, 065017 (2016),
% doi:10.1103/PhysRevD.93.065017
arXiv:1511.00933 [astro-ph.HE].

\bibitem{Blandford:2018iot}
R.~Blandford, D.~Meier, and A.~Readhead,
%``Relativistic Jets from Active Galactic Nuclei,''
Ann. Rev. Astron. Astrophys. \textbf{57}, 467 (2019),
%doi:10.1146/annurev-astro-081817-051948
arXiv:1812.06025 [astro-ph.HE].

\bibitem{Beskin:2010iba} 
V.~S.~Beskin, 
{\sl MHD Flows in Compact Astrophysical Objects: Accretion, Winds and Jets} (Springer, Heidelberg, 2010).
%doi:10.1007/978-3-642-01290-7

\bibitem{Turolla:2015mwa}
R.~Turolla, S.~Zane, and A.~Watts,
%``Magnetars: the physics behind observations. A review,''
Rept. Prog. Phys. \textbf{78}, 116901 (2015),
%doi:10.1088/0034-4885/78/11/116901
arXiv:1507.02924 [astro-ph.HE].

\bibitem{Kaspi:2017fwg}
V.~M.~Kaspi and A.~Beloborodov,
%``Magnetars,''
Ann. Rev. Astron. Astrophys. \textbf{55}, 261 (2017),
%doi:10.1146/annurev-astro-081915-023329
arXiv:1703.00068 [astro-ph.HE].

\bibitem{Gourgouliatos:2018efn}
K.~N.~Gourgouliatos and P.~Esposito,
%``Strongly magnetized pulsars: explosive events and evolution,''
Astrophys. Space Sci. Libr. \textbf{457}, 57 (2018),
%doi:10.1007/978-3-319-97616-7\_2
arXiv:1805.01680 [astro-ph.HE].

\bibitem{vandenEijnden:2018moe}
J.~van den Eijnden, N.~Degenaar, T.~D.~Russell, R.~Wijnands, J.~C.~A.~Miller-Jones, G.~R.~Sivakoff, and J.~V.~Hern\'andez Santisteban,
%``An evolving jet from a strongly magnetized accreting X-ray pulsar,''
Nature \textbf{562}, 233 (2018),
%doi:10.1038/s41586-018-0524-1
arXiv:1809.10204 [astro-ph.HE].

\bibitem{Nishikawa:2020rwe}
K.~Nishikawa, I.~Dutan, C.~Koehn, and Y.~Mizuno,
%``PIC methods in astrophysics: Simulations of relativistic jets and kinetic physics in astrophysical systems,''
Liv. Rev. Comput. Astrophys. \textbf{7}, 1 (2021),
%doi:10.1007/s41115-021-00012-0
arXiv:2008.02105 [astro-ph.HE].

\bibitem{Komissarov:2021}  
S.~Komissarov and O.~Porth,
%``Numerical simulations of jets,"
New Astron. Rev. \textbf{92}, 101610 (2021).
%https://doi.org/10.1016/j.newar.2021.101610.

\bibitem{Sturrock:1971zc}
P.~A.~Sturrock,
%``A Model of pulsars,''
Astrophys. J. \textbf{164}, 529 (1971).
%doi:10.1086/150865

\bibitem{Ruderman:1975ju}
M.~A.~Ruderman and P.~G.~Sutherland,
%``Theory of pulsars: Polar caps, sparks, and coherent microwave radiation,''
Astrophys. J. \textbf{196}, 51 (1975).
%doi:10.1086/153393

\bibitem{Arons:1983aa} J.~Arons, 
%``Pair creation above pulsar polar caps : geometrical structure and energetics of slot gaps,"
Astrophys. J. \textbf{266}, 215 (1983). 
%doi:10.1086/160771

\bibitem{Cheng:1986qt}
K.~S.~Cheng, C.~Ho, and M.~A.~Ruderman,
%``Energetic Radiation from Rapidly Spinning Pulsars. 1. Outer Magnetosphere Gaps. 2. Vela and Crab,''
Astrophys. J. \textbf{300}, 500 (1986).
%doi:10.1086/163829

\bibitem{Prabhu:2021zve}
A.~Prabhu,
%``Axion production in pulsar magnetosphere gaps,''
Phys. Rev. D \textbf{104}, 055038 (2021),
%doi:10.1103/PhysRevD.104.055038
arXiv:2104.14569 [hep-ph].

\bibitem{Masui:2015kmb}
K.~Masui %H.~H.~Lin, J.~Sievers, C.~J.~Anderson, T.~C.~Chang, X.~Chen, A.~Ganguly, M.~Jarvis, C.~Y.~Kuo and Y.~C.~Li, 
\textit{et al.},
%``Dense magnetized plasma associated with a fast radio burst,''
Nature \textbf{528}, 523 (2015),
%doi:10.1038/nature15769
arXiv:1512.00529 [astro-ph.HE].

\bibitem{Champion:2015pmj}
D.~J.~Champion % E.~Petroff, M.~Kramer, M.~J.~Keith, M.~Bailes, E.~D.~Barr, S.~D.~Bates, N.~D.~R.~Bhat, M.~Burgay, and S.~Burke-Spolaor, 
\textit{et al.},
%``Five new fast radio bursts from the HTRU high-latitude survey at Parkes: first evidence for two-component bursts,''
Mon. Not. Roy. Astron. Soc. \textbf{460}, L30 (2016),
%doi:10.1093/mnrasl/slw069
arXiv:1511.07746 [astro-ph.HE].

\bibitem{Kulkarni:2015sxv}
S.~R.~Kulkarni, E.~O.~Ofek, and J.~D.~Neill,
%``The Arecibo Fast Radio Burst: Dense Circum-burst Medium,''
arXiv:1511.09137 [astro-ph.HE].

\bibitem{Popov:2018hkz}
S.~B.~Popov, K.~A.~Postnov, and M.~S.~Pshirkov,
%``Fast Radio Bursts,''
Phys. Usp. \textbf{61},  965 (2018),
%doi:10.3367/UFNe.2018.03.038313
arXiv:1806.03628 [astro-ph.HE].

\bibitem{Lyutikov:2002kh}
M.~Lyutikov,
%``Radio emission from magnetars,''
Astrophys. J. Lett. \textbf{580}, L65 (2002),
%doi:10.1086/345493
arXiv:astro-ph/0206439 [astro-ph].

\bibitem{CHIMEFRB:2020abu}
B.~C.~Andersen \textit{et al.} [CHIME/FRB],
%``A bright millisecond-duration radio burst from a Galactic magnetar,''
Nature \textbf{587}, 54 (2020),
%doi:10.1038/s41586-020-2863-y
arXiv:2005.10324 [astro-ph.HE].

\bibitem{Bochenek:2020zxn}
C.~D.~Bochenek, V.~Ravi, K.~V.~Belov, G.~Hallinan, J.~Kocz, S.~R.~Kulkarni, and D.~L.~McKenna,
%``A fast radio burst associated with a Galactic magnetar,''
Nature \textbf{587}, 59 (2020),
%doi:10.1038/s41586-020-2872-x
arXiv:2005.10828 [astro-ph.HE].

\bibitem{Mereghetti:2020unm}
S.~Mereghetti % V.~Savchenko, C.~Ferrigno, D.~G\"otz, M.~Rigoselli, A.~Tiengo, A.~Bazzano, E.~Bozzo, A.~Coleiro and T.~J.~L.~Courvoisier, 
\textit{et al.},
%``INTEGRAL discovery of a burst with associated radio emission from the magnetar SGR 1935+2154,''
Astrophys. J. Lett. \textbf{898}, L29 (2020),
%doi:10.3847/2041-8213/aba2cf
arXiv:2005.06335 [astro-ph.HE].

\bibitem{Tavani:2020ATe} M.~Tavani % A.~Ursi, F.~Verrecchia, 
\textit{et al.},
%``AGILE detection of a hard X-ray burst in temporal coincidence with a radio burst from SGR 1935+2154,"
Astronomer’s Telegram 13686 (2020).

\bibitem{Ridnaia:2020gcv}
A.~Ridnaia
%D.~Svinkin, D.~Frederiks, A.~Bykov, S.~Popov, R.~Aptekar, S.~Golenetskii, A.~Lysenko, A.~Tsvetkova, and M.~Ulanov,
\textit{et al.},
%``A peculiar hard X-ray counterpart of a Galactic fast radio burst,''
Nature Astron. \textbf{5},  372 (2021),
%doi:10.1038/s41550-020-01265-0
arXiv:2005.11178 [astro-ph.HE].

\bibitem{Insight-HXMTTeam:2020dmu}
C.~K.~Li \textit{et al.} [Insight-HXMT Team],
%``Identification of a non-thermal X-ray burst with the Galactic magnetar SGR 1935+2154 and a fast radio burst with Insight-HXMT,''
Nature Astron. \textbf{5}, 378 (2021),
%doi:10.1038/s41550-021-01302-6
arXiv:2005.11071 [astro-ph.HE].

\bibitem{Lyutikov:2019bcu}
M.~Lyutikov and M.~Rafat,
%``Coherence constraints on physical parameters at bright radio sources and FRB emission mechanism,''
arXiv:1901.03260 [astro-ph.HE].

\bibitem{Philippov:2020jxu}
A.~Philippov, A.~Timokhin, and A.~Spitkovsky,
%``Origin of Pulsar Radio Emission,''
Phys. Rev. Lett. \textbf{124}, 245101 (2020),
%doi:10.1103/PhysRevLett.124.245101
arXiv:2001.02236 [astro-ph.HE].

\bibitem{Landau:1932} L.~D.~Landau, 
%``On the theory of stars,''
Phys. Z. Sowjetunion \textbf{1}, 285 (1932).

\bibitem{Baade:1934zex}
W.~Baade and F.~Zwicky,
%``On Super-Novae,''
Proc. Nat. Acad. Sci. \textbf{20}, 254 (1934).
%doi:10.1073/pnas.20.5.254

\bibitem{Hewish:1968bj}
A.~Hewish, S.~J.~Bell, J.~D.~H.~Pilkington, P.~F.~Scott, and R.~A.~Collins,
%``Observation of a rapidly pulsating radio source,''
Nature \textbf{217}, 709 (1968).
%I doi:10.1038/217709a0

\bibitem{Melrose:2016kaf}
D.~B.~Melrose and R.~Yuen,
%``Pulsar Electrodynamics: an unsolved problem,''
J. Plasma Phys. \textbf{82},  635820202 (2016),
%doi:10.1017/S0022377816000398
arXiv:1604.03623 [astro-ph.HE].

\bibitem{Goldreich:1969sb}
P.~Goldreich and W.~H.~Julian,
%``Pulsar electrodynamics,''
Astrophys. J. \textbf{157}, 869 (1969).
%doi:10.1086/150119

\bibitem{Medin:2007vd}
Z.~Medin and D.~Lai,
%``Condensed Surfaces of Magnetic Neutron Stars, Thermal Surface Emission, and Particle Acceleration Above Pulsar Polar Caps,''
AIP Conf. Proc. \textbf{983}, 249 (2008),
%doi:10.1063/1.2900154
arXiv:0708.3863 [astro-ph].

\bibitem{Petri:2016tqe} 
J.~P\'etri,
%``Theory of pulsar magnetosphere and wind,''
J. Plasma Phys. \textbf{82}, 635820502 (2016)
%doi:10.1017/S0022377816000763
arXiv:1608.04895 [astro-ph.HE].

\bibitem{Lieu:2016hfw}
R.~Lieu,
%``Are fast radio bursts the birthmark of magnetars?,''
Astrophys. J. \textbf{834}, 199 (2017),
%doi:10.3847/1538-4357/834/2/199
arXiv:1611.03094 [astro-ph.HE].

\bibitem{Adler:1969gk}
S.~L.~Adler,
%``Axial vector vertex in spinor electrodynamics,''
Phys. Rev. \textbf{177}, 2426 (1969).
%doi:10.1103/PhysRev.177.2426

\bibitem{Bell:1969ts}
J.~S.~Bell and R.~Jackiw,
%``A PCAC puzzle: $\pi^0 \to \gamma \gamma$ in the $\sigma$ model,''
Nuovo Cim. A \textbf{60}, 47 (1969).
%doi:10.1007/BF02823296

\bibitem{Boyarsky:2020cyk}
A.~Boyarsky, V.~Cheianov, O.~Ruchayskiy, and O.~Sobol,
%``Evolution of the Primordial Axial Charge across Cosmic Times,''
Phys. Rev. Lett. \textbf{126}, 021801 (2021),
%doi:10.1103/PhysRevLett.126.021801
arXiv:2007.13691 [hep-ph].

\bibitem{Boyarsky:2020ani}
A.~Boyarsky, V.~Cheianov, O.~Ruchayskiy, and O.~Sobol,
%``Equilibration of the chiral asymmetry due to finite electron mass in electron-positron plasma,''
Phys. Rev. D \textbf{103}, 013003 (2021),
%doi:10.1103/PhysRevD.103.013003
arXiv:2008.00360 [hep-ph].

\bibitem{Copinger:2018ftr}
P.~Copinger, K.~Fukushima, and S.~Pu,
%``Axial Ward identity and the Schwinger mechanism -- Applications to the real-time chiral magnetic effect and condensates,''
Phys. Rev. Lett. \textbf{121}, 261602 (2018),
%doi:10.1103/PhysRevLett.121.261602
arXiv:1807.04416 [hep-th].

\bibitem{Grabowska:2014efa}
D.~Grabowska, D.~B.~Kaplan, and S.~Reddy,
%``Role of the electron mass in damping chiral plasma instability in Supernovae and neutron stars,''
Phys. Rev. D \textbf{91},  085035 (2015),
%doi:10.1103/PhysRevD.91.085035
arXiv:1409.3602 [hep-ph].

\bibitem{East:2021spd}
W.~E.~East, L.~Lehner, S.~L.~Liebling, and C.~Palenzuela,
%``Multimessenger Signals from Black Hole\textendash{}Neutron Star Mergers without Significant Tidal Disruption,''
Astrophys. J. Lett. \textbf{912}, L18 (2021),
%doi:10.3847/2041-8213/abf566
arXiv:2101.12214 [astro-ph.HE].

\bibitem{Eatough:2013nva}
R.~P.~Eatough \textit{et al.}, % H.~Falcke, R.~Karuppusamy, K.~J.~Lee, D.~J.~Champion, E.~F.~Keane, G.~Desvignes, D.~H.~F.~M.~Schnitzeler, L.~G.~Spitler and M.~Kramer, \textit{et al.}
%``A strong magnetic field around the supermassive black hole at the centre of the Galaxy,''
Nature \textbf{501}, 391 (2013),
%doi:10.1038/nature12499
arXiv:1308.3147 [astro-ph.GA].

\bibitem{EventHorizonTelescope:2021srq}
K.~Akiyama \textit{et al.} [Event Horizon Telescope],
%``First M87 Event Horizon Telescope Results. VIII. Magnetic Field Structure near The Event Horizon,''
Astrophys. J. Lett. \textbf{910}, L13 (2021)
%doi:10.3847/2041-8213/abe4de
arXiv:2105.01173 [astro-ph.HE].

\bibitem{EventHorizonTelescope:2019pgp}
K.~Akiyama \textit{et al.} [Event Horizon Telescope],
%``First M87 Event Horizon Telescope Results. V. Physical Origin of the Asymmetric Ring,''
Astrophys. J. Lett. \textbf{875}, L5 (2019),
%doi:10.3847/2041-8213/ab0f43
arXiv:1906.11242 [astro-ph.GA].

\bibitem{Beskin:1992aaa} V.~S.~Beskin, Y.~N.~Istomin, and V.~I.~Parev, 
Sov. Astron. Lett. \textbf{36}, 642 (1992).

\bibitem{Ptitsyna:2015nta}
K.~Ptitsyna and A.~Neronov,
%``Particle acceleration in the vacuum gaps in black hole magnetospheres,''
Astron. Astrophys. \textbf{593}, A8 (2016),
%doi:10.1051/0004-6361/201527549
arXiv:1510.04023 [astro-ph.HE].

\bibitem{Arendt:2002ay}
P.~N.~Arendt, Jr. and J.~A.~Eilek,
%``Pair creation in the pulsar magnetosphere,''
Astrophys. J. \textbf{581}, 451 (2002),
%doi:10.1086/344133
arXiv:astro-ph/0207638 [astro-ph].

\bibitem{Akamatsu:2013pjd} 
Y.~Akamatsu and N.~Yamamoto,
%``Chiral Plasma Instabilities,''
Phys. Rev. Lett. \textbf{111}, 052002 (2013),
%doi:10.1103/PhysRevLett.111.052002
arXiv:1302.2125 [nucl-th].

\bibitem{Arnold:2003zc}
P.~B.~Arnold, G.~D.~Moore, and L.~G.~Yaffe,
%``Transport coefficients in high temperature gauge theories. 1. Leading log results,''
JHEP \textbf{05}, 051 (2003),
%doi:10.1088/1126-6708/2003/05/051
arXiv:hep-ph/0302165 [hep-ph].

\bibitem{Chandrasekhar:1957} 
S.~Chandrasekhar and P.~C.~Kendall, 
%``On Force-Free Magnetic Fields,"
%doi:  10.1086/146413
Astrophys. J. \textbf{126}, 457 (1957).

\bibitem{Rogachevskii:2017uyc}
I.~Rogachevskii, O.~Ruchayskiy, A.~Boyarsky, J.~Fr\"ohlich, N.~Kleeorin, A.~Brandenburg, and J.~Schober,
%``Laminar and turbulent dynamos in chiral magnetohydrodynamics-I: Theory,''
Astrophys. J. \textbf{846}, 153 (2017),
%doi:10.3847/1538-4357/aa886b
arXiv:1705.00378 [physics.plasm-ph].

\bibitem{Brandenburg:2017rcb}
A.~Brandenburg, J.~Schober, I.~Rogachevskii, T.~Kahniashvili, A.~Boyarsky, J.~Frohlich, O.~Ruchayskiy, and N.~Kleeorin,
%``The turbulent chiral-magnetic cascade in the early universe,''
Astrophys. J. Lett. \textbf{845},  L21 (2017),
%doi:10.3847/2041-8213/aa855d
arXiv:1707.03385 [astro-ph.CO].

\bibitem{Schober:2017cdw}
J.~Schober, I.~Rogachevskii, A.~Brandenburg, A.~Boyarsky, J.~Fr\"ohlich, O.~Ruchayskiy, and N.~Kleeorin,
%``Laminar and turbulent dynamos in chiral magnetohydrodynamics. II. Simulations,''
Astrophys. J. \textbf{858}, 124 (2018)
%doi:10.3847/1538-4357/aaba75
arXiv:1711.09733 [physics.flu-dyn].


\end{thebibliography}
\end{document}